\documentclass[aps,prd,showpacs,nobibnotes,twocolumn,preprintnumbers,
nobalancelastpage,superscriptaddress]{revtex4}

\usepackage{graphicx}
\usepackage{amsmath}
\usepackage[svgnames]{xcolor} %OMG - use SVG, not DVI or ODP or PS or POOP
\usepackage[colorlinks=True, linkcolor=SlateBlue,
citecolor=SteelBlue,urlcolor=BlueViolet]{hyperref}
\usepackage[capitalise]{cleveref}

\usepackage{aas_macros}

\begin{document}
	\title{ A linear relation between galaxy-lensing cross-correlations to test the cosmological principle model-independently}
	
	\author{Hai Yu}
	\email[]{yu\_hai@sjtu.edu.cn}
	\affiliation{Department of Astronomy, School of Physics and Astronomy, Shanghai Jiao Tong University, Shanghai, 200240, China}
	\affiliation{Key Laboratory for Particle Astrophysics and Cosmology (MOE)/Shanghai Key Laboratory for Particle Physics and Cosmology, China}
	\author{Pengjie Zhang}
	\email[]{zhangpj@sjtu.edu.cn}
	\affiliation{Department of Astronomy, School of Physics and Astronomy, Shanghai Jiao Tong University, Shanghai, 200240, China}
	\affiliation{Key Laboratory for Particle Astrophysics and Cosmology (MOE)/Shanghai Key Laboratory for Particle Physics and Cosmology, China}
	\affiliation{Tsung-Dao Lee Institute, Shanghai Jiao Tong University, Shanghai, 200240, China}
	
	\author{Jiaxin Wang}
	\affiliation{Department of Astronomy, School of Physics and Astronomy, Shanghai Jiao Tong University, Shanghai, 200240, China}
	\affiliation{Key Laboratory for Particle Astrophysics and Cosmology (MOE)/Shanghai Key Laboratory for Particle Physics and Cosmology, China}
	\author{Ji Yao}
	\affiliation{Department of Astronomy, School of Physics and Astronomy, Shanghai Jiao Tong University, Shanghai, 200240, China}
	\affiliation{Key Laboratory for Particle Astrophysics and Cosmology (MOE)/Shanghai Key Laboratory for Particle Physics and Cosmology, China}
	\author{Fa-Yin Wang}
	\affiliation{School of Astronomy and Space Science, Nanjing University, Nanjing 210093, China}
	\affiliation{Key Laboratory of Modern Astronomy and Astrophysics (Nanjing University), Ministry of Education, Nanjing 210093, China}
	
	%\date{\today}
	
	\begin{abstract}
		We discover a linear relation between two sets of galaxy-lensing cross-correlations. This linear relation holds, as long as light follows the geodesic and the metric is Friedmann-Lema\^{i}tre-Robertson-Walker (FLRW).  Violation of the cosmological principle (and equivalently the FLRW metric) will break this linear relation. Therefore it provides a powerful test of the cosmological principle, based on direct observables and relied on no specific cosmological models. We demonstrate that stage IV galaxy surveys and CMB-S4 experiments will be able to test this linear relation stringently and therefore test the cosmological principle robustly. 
	\end{abstract}
	\maketitle
	{\bf Introduction}.---
	The $\Lambda$CDM model has achieved splendid success in describing the observed universe (e.g. \cite{2020A&A...641A...1P}). This standard paradigm of modern cosmology is based upon the cosmological principle. It claims that our universe follows statistical homogeneity and isotropy on large scales\cite{Weinberg1972gcpa.book.....W}. 
	Nevertheless, robust verification of the cosmological principle (or equivalently the Friedmann-Lema\^{i}tre-Robertson-Walker (FLRW) metric) still remains a daunting task of modern cosmology.   Primary CMB observation (e.g. \cite{PlanckCollaboration2020A&A...641A...6P}) verifies the isotropy of the universe to a high confidence level. However, this verification is restricted to the view of the universe from a specific location (the solar system). It lacks the power to constrain inhomogeneous but spherically symmetric models such as the Lema\^{i}tre-Tolmen-Bondi(LTB) model \cite{1947MNRAS.107..410B}.  Such radial inhomogeneity (violation of the Copernican principle) can be tested with large scale distribution of galaxies \cite{Hogg2005ApJ...624...54H,Sarkar2009MNRAS.399L.128S} and quasars \cite{Goncalves2018MNRAS.481.5270G,2021JCAP...03..029G},  but the evolution of the universe along the radial direction has to be appropriately corrected. Type Ia supernovae have also been used for such tests (e.g. test of the void models which can replace dark energy\cite{2008PhRvL.101m1302C,2008PhRvL.101y1303Z}).
	
	Meanwhile, several tests independent of assumptions such as dark energy have been proposed. The radial inhomogeneity is often parametrized as an effective curvature varying in redshift. It can be measured independent of specific cosmological models, through consistency relations between distances, $H(z)$, or redshift drift  \cite{Uzan2008PhRvL.100s1303U,Clarkson2008PhRvL.101a1301C,Rasanen2015PhRvL.115j1301R}. However, it requires multiple data sets such as angular diameter/luminosity distance and/or Hubble parameter $H(z)$ (e.g. \cite{2016ApJ...833..240L,2020ApJ...898..100W,2020ApJ...901..129L}). Furthermore, some of these data are derived under cosmological (e.g. modeling of baryon acoustic oscillation \cite{2021MNRAS.500..736B}) or astrophysical (e.g. modeling of lens \cite{2017ApJ...834...75X}) assumptions. Secondary CMB anisotropies provide alternative tests \cite{1995PhRvD..52.1821G,2007astro.ph..3541S}, using free electrons as distant observers. The dipole on their CMB last scattering surfaces generates the kinetic Sunyaev-Zel’dovich (kSZ) effect \cite{Zhang2011PhRvL.107d1301Z,2015JCAP...06..046Z}, and the quadrupole generates spectral distortion \cite{2008PhRvL.100s1302C}.  These tests have ruled out strong violation of the Copernican principle \cite{Zhang2011PhRvL.107d1301Z}, yet further improvement awaits precision measurement of the kSZ effect and spectral distortions. 
	
	Here we propose a new test of the cosmological principle, based on direct weak lensing observables that will be measured to high accuracy by upcoming surveys. It extends the idea of isolating geometry with galaxy-lensing tomography\cite{Jain2003PhRvL..91n1302J,Zhang2005ApJ...635..806Z}.  We find a linear relation between two sets of galaxy-lensing cross-correlations. The two sets have different lens(foreground) galaxy distributions,  but share the same source(background) distribution. With fixed lens distributions and varying source distributions, the two sets form a linear relation in an FLRW universe. Here the lenses offer a valuable opportunity to view the universe from a distant location. When the universe viewed from one lens differs from the universe viewed from the other lens or the solar system,  the linear relation breaks. We are then able to justify/falsify the  FLRW metric (or equivalently the cosmological principle). This test makes no assumptions on cosmological models and is therefore free to complexities in dark energy or modified gravity. 
	
	{\bf The linear relation}.---
	Distribution of foreground galaxies and weak lensing of background galaxies and diffuse cosmic backgrounds (e.g. CMB) are spatially correlated. We can measure such cross-correlation combining galaxies with spectroscopic/photometric redshifts, and weak lensing obtained from cosmic shear, cosmic magnification or CMB lensing. Under the Limber approximations \cite{Limber1954ApJ...119..655L,LoVerde2008PhRvD..78l3506L}, the cross-power spectrum is \cite{Hu2004PhRvD..70d3009H}
	\begin{equation}\label{eq:C_rg}
	C_\ell= \int_{0}^{\infty} d\chi \frac{n_{\mathrm{f}}(\chi)q(\chi)\ell^2}{f_\mathrm{K}^2 }P_{\mathrm{g}\Phi}\left(k=\frac{\ell}{f_\mathrm{K}(\chi)},\chi\right).
	\end{equation}
	Here $\chi$ is the comoving radial coordinate. $f_\mathrm{K}(\chi)$ is the comoving angular diameter distance. $n_{\mathrm{f}}(\chi)$ is the normalized distribution of the foreground galaxies and $P_{\mathrm{g}\Phi}(k,\chi)$ is the three-dimensional cross-power spectrum between galaxy number density and gravitational potential. The lensing kernel is given by
	\begin{equation}\label{eq:q_j}
	q(\chi) = \int_{\chi}^{\infty}d\chi_{\mathrm{s}} n_{\mathrm{s}}(\chi_{\mathrm{s}}) B(\chi,\chi_{\mathrm{s}})\ ,
	\end{equation}
	\begin{equation}\label{eq:B}
	B(\chi_\mathrm{f},\chi_\mathrm{s}) \equiv \frac{f_\mathrm{K}(\chi_\mathrm{f},\chi_\mathrm{s})}{f_\mathrm{K}(\chi_\mathrm{f})f_\mathrm{K}(\chi_\mathrm{s})}\ .
	\end{equation}
	Here $f_\mathrm{K}(\chi_\mathrm{f},\chi_\mathrm{s})$ is the comoving angular diameter distance to $\chi_{\mathrm{s}}$, measured at $\chi_{\mathrm{f}}$. $f_{\mathrm{K}}(\chi)\equiv f_{\mathrm{K}}(0,\chi)$.  Eq. \ref{eq:C_rg} is valid as long as light follows the geodesic \cite{Jain08}. It is more general than the usual expression with the galaxy-matter power spectrum $P_{\mathrm{gm}}$, which assumes general relativity. 
	
	For sufficiently narrow distribution of lens and source galaxies, Eq.(\ref{eq:C_rg}) reduces to 
	\begin{equation}\label{eq:C_lfb}
	C_\ell(\chi_\mathrm{f},\chi_\mathrm{s}) = A(\ell,\chi_\mathrm{f})B(\chi_\mathrm{f},\chi_\mathrm{s})\ ,
	\end{equation}
	\begin{equation}\label{eq:A}
	A(\ell,\chi_\mathrm{f}) \equiv \frac{\ell^2}{f^2_\mathrm{K}(\chi_\mathrm{f})}P_{\mathrm{g}\Phi}\left(\frac{\ell}{f_\mathrm{K}(\chi_\mathrm{f})},\chi_\mathrm{f}\right)\ .
	\end{equation}
	$A(\ell,\chi_\mathrm{f})$ only depends on the foreground galaxy distribution. Therefore by fixing the foreground galaxies and varying the source galaxy distribution, weak lensing provides a clean geometry measure of the universe \cite{Jain2003PhRvL..91n1302J,Zhang2005ApJ...635..806Z, Bernstein2006ApJ...637..598B, Prat_2019, 2021PhRvD.103d3539Z}.  
	
	What we find is a new relation,  revealed by the unexplored configuration with two fixed foreground (lens) galaxy distributions at $\chi_{\mathrm{f}_{1,2}}$, and varying source galaxy distribution at $\chi_\mathrm{s}$ ($\chi_{\mathrm{f}_1}<\chi_{\mathrm{f}_2}<\chi_\mathrm{s}$). This relation reads
	\begin{equation}\label{eq:wl_relation}
	C_\ell(\chi_{\mathrm{f}_1},\chi_\mathrm{s}) = F_{12}C_\ell(\chi_{\mathrm{f}_2},\chi_\mathrm{s})+C_{12}\ .
	\end{equation}
	Here $F_{12}\equiv A(\ell,\chi_{\mathrm{f}_1})/A(\ell,\chi_{\mathrm{f}_2})$ and $C_{12}\equiv C_\ell(\chi_{\mathrm{f}_1},\chi_{\mathrm{f}_2})$. Both $F_{12}$ and $C_{12}$ are constants for fixed foreground distributions (and fixed multipole). Therefore when we only vary the source distribution, Eq. \ref{eq:wl_relation} reveals a linear relation between $y\equiv C_\ell(\chi_{\mathrm{f}_1}, \chi_\mathrm{s})$ and $x\equiv C_\ell(\chi_{\mathrm{f}_2},\chi_\mathrm{s})$,
	\begin{equation}
	y=ax+b\ .
	\end{equation}
	
	The proof of this relation is straightforward. For the FLRW metric, 
	\begin{eqnarray}\label{eq:f_K}
	f_\mathrm{K}(\chi_1, \chi_2) &=& \frac{1}{\eta}
	\begin{cases}
	\eta(\chi_2-\chi_1) & \text{if } \Omega_\mathrm{k} = 0,\\
	\sin[\eta(\chi_2-\chi_1)] & \text{if } \Omega_\mathrm{k} > 0, \\
	\sinh[\eta (\chi_2-\chi_1)] & \text{if } \Omega_\mathrm{k} < 0,
	\end{cases}
	\end{eqnarray}
	where $\eta=H_0\sqrt{|\Omega_\mathrm{k}|}$. Here $H_0$ is the Hubble constant and $\Omega_\mathrm{k}$ is the cosmological curvature parameter. Then we can verify straightforwardly the following relation
	\begin{equation}\label{eq:B_relation}
	B(\chi_{\mathrm{f}_1},\chi_\mathrm{s}) = B(\chi_{\mathrm{f}_1},\chi_{\mathrm{f}_2})+B(\chi_{\mathrm{f}_2},\chi_\mathrm{s}).
	\end{equation}
	Together with Eq. \ref{eq:C_lfb}, we prove Eq. \ref{eq:wl_relation}.  This linear relation holds under two conditions, that the cosmological principle is valid, and light travels along the geodesic. We further address that, although this linear relation is first derived assuming narrow source distribution, it actually holds for arbitrarily wide source distribution, as long as it does not overlap with the two lens distributions. 
	
	The two variables $C_\ell(\chi_{\mathrm{f}_1},\chi_\mathrm{s})$ and $C_\ell(\chi_{\mathrm{f}_2},\chi_\mathrm{s})$ are both direct observables, so we are able to directly test this linear relation with 3 or more source distributions.
	From the derivation, this linear relation is a necessary condition for the FLRW metric. Therefore it enables a direct test of the FLRW metric, based purely on observables.  Since the FLRW metric is equivalent to the cosmological principle, it is also a direct test of the cosmological principle. Namely,  the failure of the linear relation will falsify the cosmological principle and lead to the discovery of a non-Copernican universe. The test does not require priors on cosmological models (e.g. dark energy/modified gravity) and/or cosmological parameters, other than that light follows the geodesic.
	
	{\bf The constraining power}.---
	We demonstrate the constraining power of the linear relation against the violation of the cosmological principle. The cosmological principle guarantees the curvature viewed from the two locations are identical. But this will be no longer valid when the cosmological principle is violated. Therefore violation of the cosmological principle is often parametrized with an effective curvature of the universe varying with redshift and direction (e.g. \cite{Clarkson2008PhRvL.101a1301C,Rasanen2015PhRvL.115j1301R}).  For our study, Eq. \ref{eq:B} \& \ref{eq:f_K} tell us that the lensing kernel is determined by both the angular diameter distances viewed from the solar system, and that viewed from the lens. Therefore we parametrize violation of the cosmological principle by three effective curvatures, as viewed from the solar system ($\Omega_{K,0}$), and from the two lenses ($\Omega_{K,1}$ and $\Omega_{K,2}$).  When any of the two or all of them differ,  the cosmological principle is violated and the linear relation is broken. For brevity, we fix other cosmological parameters as the Planck18 best-fit values\cite{2020A&A...641A...1P}. We also fix $\Omega_{K,0}=0$ for brevity.	
	
	%%%%%%%%%%%%%%%%%%%%%%%%%%%%%%%%%
	\begin{figure}
		\centering
		\includegraphics[width=0.47\textwidth]{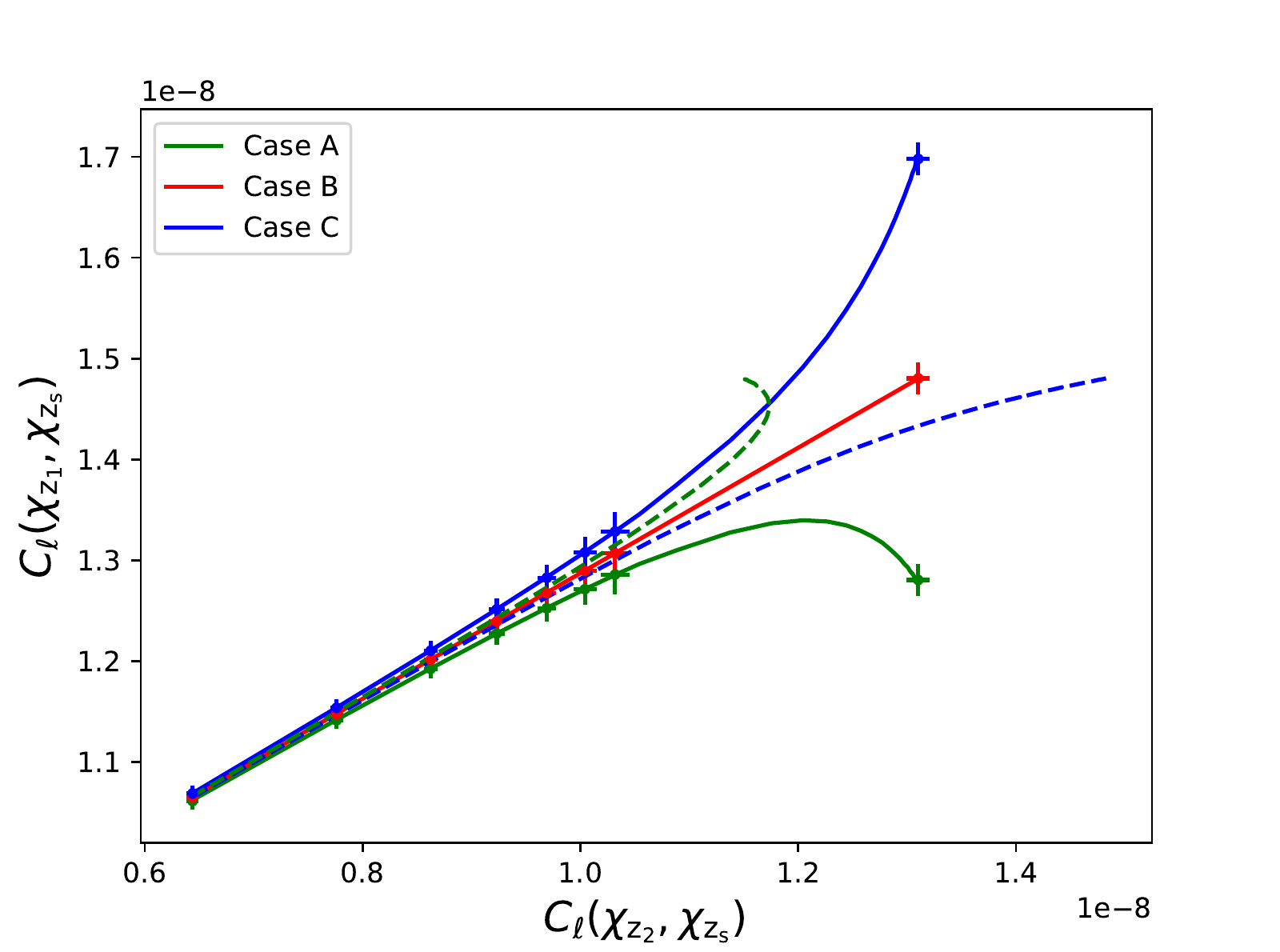}\\
		\caption{The linear relation in the FLRW metric and its violation in non-FLRW universes, in which the cosmological principle is violated. The x-axis and y-axis are both direct observables, the galaxy-lensing cross-correlations. The two lens distributions are fixed, but the source distribution spans from low redshift (shear) to the last scattering surface (CMB lensing, the last data points).   Error estimations are for stage IV dark energy surveys such as LSST and CMB-S4. The case B corresponds to the FLRW metric while A and C are non-FLRW metrics. Case B numerically verifies the linear relation in the FLRW metric, and case A/C shows that this relation can be significantly violated in non-FLRW metrics. The two unlabelled dashed lines are results of other non-FLRW metrics, which demonstrate that violation of the linear relation is generic in non-FLRW metrics. For brevity, we only show the case of $\ell\in[700,1200]$, two lens bins at $[0.1,0.3]$ and $[0.3,0.5]$ and source bins at $0.7<z<2.1$ and $z\sim 1100$. Results for other multipole bins and lens-source redshift configurations are similar. More details are given in the main text.  }\label{fig1}
	\end{figure}
	%%%%%%%%%%%%%%%%%%%%%%%%%
	
	Fig. \ref{fig1} shows three examples for  lens distributions at $z_1\in[0.1,0.3]$ and $z_2\in[0.3,0.5]$. Case B corresponds to the FLRW metric ($(\Omega_{K,1},\Omega_{K,2})=(0,0)$). Case A has $(\Omega_{K,1},\Omega_{K,2})=(-0.1,0)$ and case C has $(\Omega_{K,1},\Omega_{K,2})=(0.1,0)$. The source distributions are $[0.5+0.2j, 0.7+0.2j]$ ($j=1,2\cdots 7$). We do not consider $z\in[0.5,0.7]$ as the source bin to avoid potential overlaping in redshift caused by photo-z errors. Furthermore, we also use CMB lensing as the most distant source at $z\sim 1100$. It is clear that both case A and case C show significant deviation from the liner relation, although the deviation from the $\Lambda$CDM prediction is within $\sim$ 10\% and is still viable with existing observations. Such violation persists for other configurations of $(\Omega_{K,1},\Omega_{K,2})$, as demonstrated by the other two unlabelled curves ($(\Omega_{K,1},\Omega_{K,2})=(0.0,\pm 0.1)$. 
	
	To estimate the errors in the cross-correlation measurement, we consider a galaxy sample with redshift distribution $n(\chi)=A\chi^\alpha\exp(-(\chi/\chi_0)^\beta)$, where $\alpha=1.0$, $\beta=4.0$ and $\chi_0=c/H_0$ \cite{Kaiser1992ApJ...388..272K,Hu1999ApJ...522L..21H}. The parameter $A$ is a normalization factor that satisfies the number density of galaxies $\int n(\chi)d\chi=20 \,\rm arcmin^{-2}$. This value of number density  is expected for the upcoming surveys such as CSST, Euclid, LSST, and WFIRST \cite{LSST2009arXiv0912.0201L,Laureijs2011arXiv1110.3193L,Schaan2017PhRvD..95l3512S,Gong2019ApJ...883..203G}. We adopt photo-z error $\sigma_{z_p} = \sigma_0(1+z)$ with $\sigma_0=0.05$, so we do not use the bin $z\in[0.5,0.7]$ as the source bin to avoid overlap between lens and source galaxy distribution. For the statistical error estimation, we adopt $f_{\rm sky}=0.3$ and galaxy shape error of $\sigma_\epsilon=0.3$ for shear measurements. For CMB lensing map, since CMB-S4\cite{2016arXiv161002743A} will reach the cosmic variance limit in the lensing reconstruction at $L\lesssim 1000$, we will take this limit for brevity. 
	
\begin{figure}
		\centering
		\includegraphics[width=0.47\textwidth]{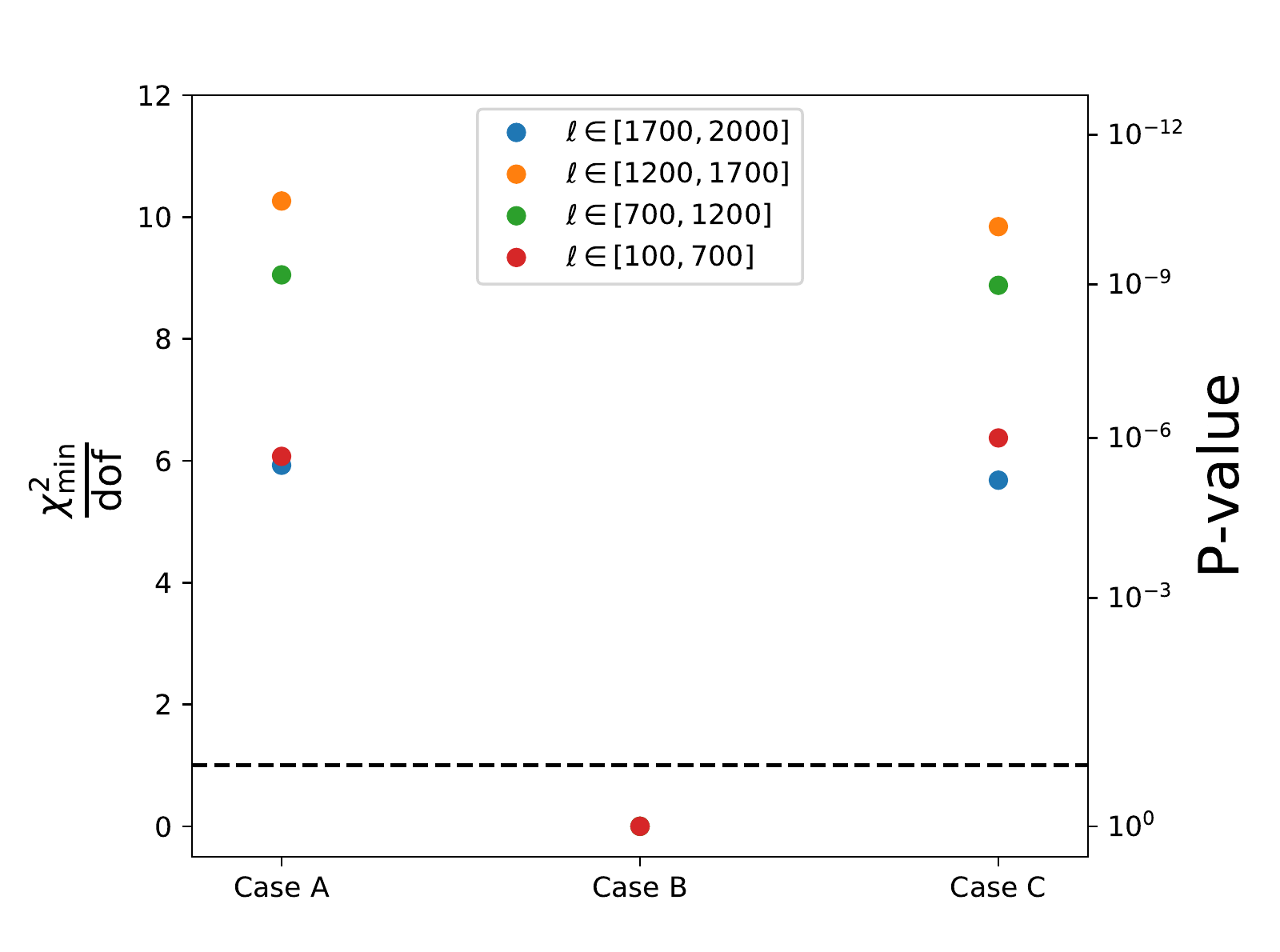}\\
		\caption{Testing the linear relation with $\chi^2_{\rm min}/{\rm dof}$ of the linear fitting. Case B (FLRW) has $\chi_{\rm min}\simeq 0$ for all 4 multipole bins investigated. This numerically verifies that the linear relation holds in the FLRW metric and shows that lens distribution with non-zero width has a negligible impact on the linear relation. In contrast, non-FLRW universes (case A/C) can not be fitted by a linear relation. The linear relation can be falsified either by a single multipole bin or a combination of several bins. Combining all multipole bins, the p-value is $10^{-31}$ for case A and $4\times 10^{-31}$ for case C, meaning that in both cases the linear relation can be robustly falsified and violation of the cosmological principle will be robustly justified. }\label{fig2}
	\end{figure}

	Fig. \ref{fig1} only shows the multipole bin $\ell\in[700,1200]$.  But this bin alone is already able to test the cosmological principle robustly. Fig. \ref{fig2} quantifies the deviation from the linear relation, with $\chi^2_{\rm min}$ of fitting the data with Eq. \ref{eq:wl_relation}. The fitting treats $F_{12}$ and $C_{12}$ as free parameters, so  the degrees of freedom is ${\rm dof}=8-2=6$. $\chi^2_{\rm min}/$dof is about $9$ for both case A and case C (Fig. \ref{fig2}). The corresponding $p$-value, namely the maximum probability of a FLRW universe to produce such data set,  is about $10^{-9}$.  

	%%%%%%%%%%%%%%%%%%%
	%%%%%%%%%%%%%%%%%%	

	The deviation from the linear relation is also significant in other multipole bins. Fig. \ref{fig2} shows the corresponding $\chi^2_{\rm min}/$dof and $p$-value for multipole bins $[100,700]$, $[700,1200]$, $[1200,1700]$ and $[1700,2000]$. Combining all the data, the probability that cases A/C is consistent with the linear relation is extremely tiny, with $p$-value $10^{-31}$ for case A and $4\times 10^{-31}$ for case C.  Therefore if such data sets are observed, the FLRW universe (and hence the cosmological principle) will be falsified to a high confidence level. 
	
\begin{figure}
		\centering
		\includegraphics[width=0.47\textwidth]{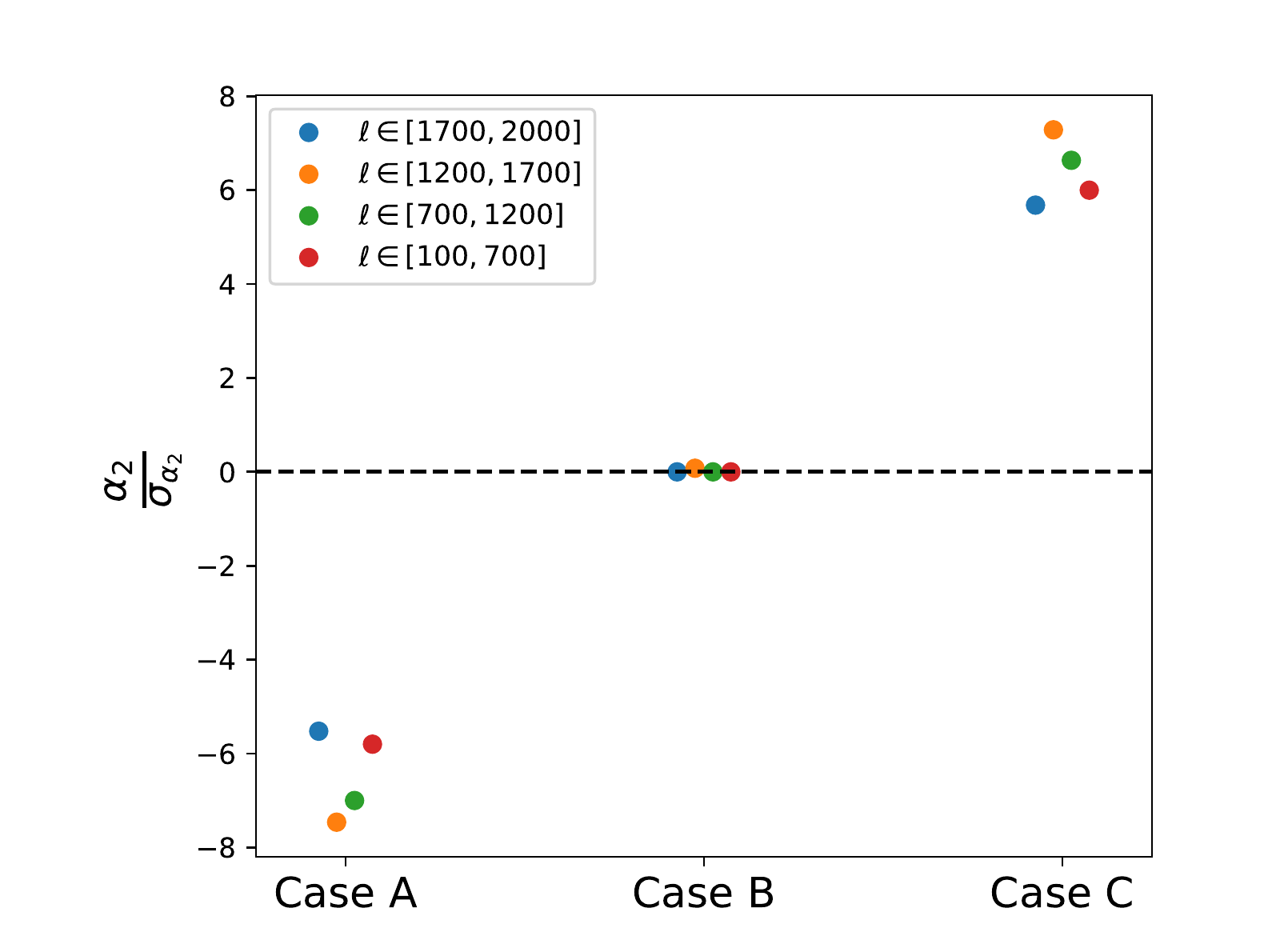}\\
		\caption{Quantifying the possible deviation of the linear relation with the quadratic fitting. Besides the $\chi^2_{\rm min}$ statistics shown in Fig. \ref{fig2}, we can quantify the deviation from the linear relation  with the detection of non-zero $\alpha_2$. Namely we fit the data points in Fig. \ref{fig1} with $y=ax+b+\alpha_2 x^2$ and check if $\alpha_2$ is consistent with zero. Case A/C have non-zero $\alpha_2$ detected at at least $5 \sigma$ for every multipole bin, showing the failure of the linear relation and violation of the cosmological principle. }\label{fig3}
	\end{figure}

	$\chi^2_{\rm min}$ of the linear fitting is not the only way of quantifying deviation from the linear relation. Another way is to fit the data points in Fig. \ref{fig1} with $y=ax+b+\alpha_2x^2$. The existence of non-zero $\alpha_2$ quantifies the deviation from the FLRW universe. Fig. \ref{fig3} shows $\alpha^{\rm bestfit}_2/\sigma_{\alpha}$ of 3 universes and 4 multipole bins. For case A/C, $\alpha^{\rm bestfit}_2$ is non-zero at at least $5 \sigma$ confidence level for every multipole bin. So the deviation from the linear relation is robustly detected. Therefore the FLRW metric can be falsified in this way as well. 
	
	Fig.\ref{fig2} \& \ref{fig3} also show $\chi^2_{\rm min}/{\rm dof}$ and  $\alpha^{\rm bestfit}_2/\sigma_{\alpha}$ for the FLRW metric (case B). As expected, they are consistent with zero.  Nevertheless, we emphasize that such checks are non-trivial. The point is that the linear relation (Eq. \ref{eq:wl_relation}) is derived under the limit of a single lens plane. In realistic surveys, the lens distribution must have a sufficiently wide bin size in order to achieve reasonable S/N in cross-correlation measurement.  Furthermore, photo-z measurements inevitably have statistical errors, which exerts a lower limit on the effective bin redshift width $\sigma_z\sim 0.05(1+z)$. Therefore the linear relation is an approximation in realistic surveys, and its accuracy must be tested. Tests on case B in Fig. \ref{fig2} \& \ref{fig3}  show that, the linear relation is nearly exact for the investigated photo-z bin width $\Delta z=0.2$ and we can safely use it to test the FLRW metric. 
	
	%%%%%%%%%%%%%%%%%%%
	%%%%%%%%%%%%%%%%%%
	
	Fig. \ref{fig1}, \ref{fig2} \& \ref{fig3} only show the test for the lens distributions at $[0.1,0.3]$ and $[0.3,0.5]$. Observationally we have other accessible lens distribution to at least $z\sim 2$, so we can carry out many more tests of the linear relation demonstrated above. This will allow us to view the universe from the eyes of lens galaxies at $0<z\lesssim 2$ and obtain a more comprehensive picture on the validity of the cosmological principle. 
	
	{\bf Discussions and conclusions.}--- 
	The linear relation test has stronger constraining power than demonstrated above. (1) So far we treat $C_{12}$ in the relation as a free parameter and ignore information in it. But actually, $C_{12}$, the galaxy-lensing between two foreground galaxy distributions, can be directly measured as well. So the constrained $C_{12}$ by fitting a linear relation can be compared against the directly measured $C_{12}$ to further test the linear relation and therefore the cosmological principle. (2) One important point learned from the demonstrated examples is that CMB lensing plays a key role in testing the linear relation, due to the large separation between source and lens/observer. For the same reason, weak lensing reconstructed from other high-redshift cosmic backgrounds such as the 21cm background and cosmic infrared background (e.g. \cite{2018PhRvD..97l3539S})  will also be highly valuable. 
	
	The linear relation has more applications than demonstrated above. (1) It can be used to test specific models such as LTB. (2) Given a parameterization of a non-FLRW metric, it can constrain the relevant parameters. For example, if we model a non-FLRW metric with the three effective curvatures, we can constraint these curvatures. But in both cases, one also needs to take the impact on matter clustering into account appropriately. This is beyond the scope of this work.  (3) It can also be applied to different patches of the sky and test the isotropy of the universe. 
	
	The linear relation test is immune to several observational effects in reality. (1) Firstly, it is insensitive to photo-z errors, since as long as the true distribution of the source galaxies locates behind that of the lens galaxies. For this reason, different source bins can overlap in redshift distributions.   Furthermore, it does not require the measurement of $C_{12}$. Therefore we can use spectroscopic galaxy samples as the lens and imaging galaxy samples as the source to further reduce potential overlap between the lens and source distribution.  (2) Secondly, it is insensitive to the intrinsic alignment in the shear measurement, as long as the lens and source distributions are non-overlapping. Magnification bias in foreground galaxies contaminates the galaxy-lensing cross-correlation.  Its main impact is to broaden the effective width of the foreground galaxy distribution. But as demonstrated with tests against case B (Fig. \ref{fig2} \& \ref{fig3}), the impact of lens bin width is negligible in testing the linear relation. Furthermore, the magnification bias induced cross-correlation is significantly smaller than the galaxy-lensing cross-correlation,  so its impact on the linear relation is further reduced.  Even better, magnification bias in background galaxy provides galaxy-lensing cross-correlation measurement (e.g. \cite{2005ApJ...633..589S}) independent of cosmic shear and CMB lensing, valuable for the linear relation test. (3) Thirdly, it is insensitive to many imaging systematics in galaxy surveys. These systematics are important for galaxy auto-correlation measurement. But they usually do not correlate with both the signal and systematics in weak lensing measurement and therefore do not bias the cross-correlation measurement. 
	
	Nevertheless, the linear relation can be contaminated by some observational effects. One is the multiplicative $m$ error in shear measurement. A constant $m$ does not alter the linear relation. But if it varies with the source distribution, it will violate the linear relation. Another is outliers in the photo-z error distribution. The true redshift of photo-z outliers can be much lower than the photo-z so that the source may locate in front of one or both lens distribution. These galaxies will have no contribution to the galaxy-lensing cross-correlation, and will therefore break the linear relation. We will investigate these issues in future works.

	{\bf Acknowledgement}.---
	HY is supported by the National Postdoctoral Program for Innovative Talents  (No. BX20190206), the Project funded by China Postdoctoral Science Foundation (No. 2019M660085), and the Super Postdoc Project of Shanghai City. PJZ, JXW, and FYW are supported by the National Science Foundation of China (11621303, 11653003, 11773021, 11890691, U1831207). JXW is also supported by the National Key R\&D Program of China (2018YFA0404504, 2018YFA0404601), the 111 project, and the CAS Interdisciplinary Innovation Team (JCTD-2019-05).

	\bibliography{ref}
	
\end{document}